\documentclass[12pt,preprint]{aastex}

\slugcomment{Accepted by ApJL on 2014 July 27}
 
\shorttitle{J1502SE/SW as Double Hotspots, not a Binary Black Hole} 

\shortauthors{Wrobel, Walker \& Fu}
\begin{document}

\title{Evidence from the Very Long Baseline Array that J1502SE/SW are
  Double Hotspots, not a Supermassive Binary Black Hole}

\author{J. M. Wrobel\altaffilmark{1,2},
  R. C. Walker\altaffilmark{1}, and H. Fu\altaffilmark{3}}

\altaffiltext{1}{National Radio Astronomy Observatory, P.O. Box O,
  Socorro, NM 87801; jwrobel@nrao.edu, cwalker@nrao.edu}

\altaffiltext{2}{The National Radio Astronomy Observatory (NRAO) is a
  facility of the National Science Foundation, operated under
  cooperative agreement by Associated Universities, Inc.}

\altaffiltext{3}{Department of Physics and Astronomy, University of
  Iowa, Van Allen Hall, Iowa City, IA 52242, USA; hai-fu@uiowa.edu}

\begin{abstract}
SDSS\,J150243.09+111557.3 is a merging system at $z = 0.39$ that hosts
two confirmed AGN, one unobscured and one dust-obscured, offset by
several kiloparsecs.  Deane et al. recently reported evidence from the
European VLBI Network (EVN) that the dust-obscured AGN exhibits two
flat-spectrum radio sources, J1502SE/SW, offset by 26 mas (140 pc),
with each source being energized by its own supermassive black hole
(BH).  This intriguing interpretation of a close binary BH was reached
after ruling out a double-hotspot scenario, wherein both hotspots are
energized by a single, central BH, a configuration occuring in the
well-studied Compact Symmetric Objects.  When observed with sufficient
sensitivity and resolution, an object with double hotspots should have
an edge-brightened structure.  We report evidence from the Very Long
Baseline Array (VLBA) for just such a structure in an image of the
obscured AGN with higher sensitivity and resolution than the EVN
images.  We thus conclude that a double-hotspot scenario should be
reconsidered as a viable interpretation for J1502SE/SW, and suggest
further VLBA tests of that scenario.  A double-hotspot scenario could
have broad implications for feedback in obscured AGNs.  We also report
a VLBA detection of high-brightness-temperature emssion from the
unobscured AGN that is offset several kiloparsecs from J1502SE/SW.
\end{abstract}


\section{Motivation}

Recent simulations of galaxy mergers are able to produce remnants that
contain two or more supermassive black holes
\citep[BHs;][]{hof07,ama10,kul12}.  When such BHs accrete, they can
appear as two or more active galactic nuclei (AGNs) on sub-galactic
scales \citep{van12,ble13}.  Systematic surveys for such multiple AGNs
can impose observational constraints on AGN activation and tidally
enhanced star formation \citep[e.g.,][]{liu12,kos12}.  Such surveys
can also constrain the BH merger rate, a key quantity for predicting
the signals expected for pulsar timing arrays and gravity-wave
detectors \citep{hob10,dot12,sha13}.  These topics are of fundamental
importance in astrophysics, so all reports of multiple AGN candidates
warrant close vetting
\citep[e.g.,][]{wan09,liu10,smi10,fu11a,fu12,com12,kos12}.

This Letter focuses on one such case, SDSS\,J150243.09+111557.3
(J1502+1115 hereafter), originally identified as a dual AGN candidate
at a redshift of $z = 0.39$ with double-peaked profiles of [OIII]
\citep{smi10} and later shown to be a merging system with an
unobscured, primary AGN, J1502P, offset by 1.4\arcsec\, (7.4 kpc)
\footnote{$H_0$ = 70 km s$^{-1}$ Mpc$^{-1}$, $\Omega_M$ = 0.3,
  $\Omega_\Lambda$ = 0.7} from a dust-obscured, secondary AGN J1502S
\citep[][F11 hereafter]{fu11b}.  \citet[][D14 hereafter]{dea14}
recently advanced J1502S as {\em itself\/} hosting two AGNs and, thus,
two supermassive BHs.  Adding in the AGN J1502P then suggests that the
merging system J1502+1115 hosts three supermassive BHs on sub-galactic
scales.

Specifically, D14 reported evidence from the European VLBI Network
(EVN) that J1502S exhibits two flat-spectrum radio sources,
J1502SE/SW, offset by 26 mas (140 pc), with each source being
energized by its own supermassive BH.  D14 reached this intriguing
interpretation of J1502SE/SW as a close binary BH after ruling out a
double-hotspot scenario, wherein both hotspots are energized by a
single, central BH in one configuration that defines the well-studied
Compact Symmetric Objects \citep[CSOs;][]{pec00}.  When observed with
sufficient sensitivity and resolution, an object with double hotspots
should have an edge-brightened structure.  This Letter reports
evidence from the Very Long Baseline Array \citep[VLBA;][]{nap94} for
just such structure in an image of J1502S with higher sensitivity and
resolution than the EVN images.  \S~\ref{imaging} presents the new
VLBA imaging and \S~\ref{implications} explores its implications.  A
summary and conclusions appear in \S~\ref{summary}.

\section{VLBA Imaging}\label{imaging}

J1502+1115 was observed during a 6-hour session with 9 VLBA antennas
on 2013 June 29 (UT) under proposal BW102 = 13A-241.  The tenth VLBA
antenna, at Fort Davis, TX, was unavailable due to equipment
malfunction.  J1504+1029 was used as a phase and relative amplitude
calibrator (Table 1).  The switching time between it and J1502+1115
was 180~s, with about a third of the session devoted to J1504+1029.
About once per hour, OQ\,208 and J1507+1236 were observed.  OQ\,208 is
a compact and slowly-varying source \citep[][and references
  therein]{wu13} observed to check the absolute flux density
calibration.  J1507+1236 (Table 1) is a VLBA calibrator with a well
known position that was observed to check the quality of the phase
referencing.

\begin{deluxetable}{llccc}
\tabletypesize{\scriptsize}
\tablecolumns{5}
\tablewidth{0pc}
\tablecaption{VLBA Astrometry and Photometry at 5 GHz}\label{tab1}
\tablehead{
\colhead{Source}    & \colhead{Peak}         & \colhead{Position}    &
\colhead{Switching} & \colhead{Integrated}\\
\colhead{ }         & \colhead{R.A./Decl.}   & \colhead{Error}       & 
\colhead{Angle}     & \colhead{Flux Density\tablenotemark{a}}\\
\colhead{ }         & \colhead{(J2000)}      & \colhead{(mas)}       & 
\colhead{(\arcdeg)} & \colhead{($\mu$Jy)}\\
\colhead{(1)}&\colhead{(2)}&\colhead{(3)}&\colhead{(4)}&\colhead{(5)}}
\startdata
J1504+1029\tablenotemark{b} & 
   15 04 24.979782 & 0.1 & \nodata & \nodata \\
 & 10 29 39.19840  & 0.1 &         & \\
J1507+1236\tablenotemark{c,d} &
   15 07 21.758063 & 0.3 & 2.23 & \nodata \\
 & 12 36 29.07573  & 0.4 &      & \\
J1502SE\tablenotemark{c} &
   15 02 43.180261 & 0.2 & 0.87 & 891$\pm$81 \\
 & 11 15 57.06831  & 0.3 &      & \\
J1502SW\tablenotemark{c} &
   15 02 43.178473 & 0.2 & 0.87 & 1010$\pm$92 \\
 & 11 15 57.06508  & 0.3 &      & \\
J1502P\tablenotemark{c}  &
   15 02 43.088667 & 0.2 & 0.87 & 254$\pm$45  \\
 & 11 15 57.40016  & 0.3 &      & \\
\enddata
\tablecomments{Col.~(1): Source.  Col.~(2): Position at peak.  Units
  of right ascension are hours, minutes, and seconds, and units of
  declination are degrees, arcminutes, and arcseconds.  Col.~(3):
  Position error.  Col.~(4): Switching angle to the phase calibrator.
  Col.~(5): Integrated flux density.  Error is the quadratic sum of
  (i) the image rms times the square root of the number of synthesized
  beam areas integrated over, (ii) a 3\% uncertainty due to
  phase-calibration errors and (iii) a 5\% uncertainty in the
  amplitide scale.}
\tablenotetext{a}{Measured from the images made from self-calibrated
  data.}
\tablenotetext{b}{Phase calibrator.  The tabulated position and errors
  are assumed and adopted from the Petrov catalog rfc\_2012b.  The
  latest Petrov catalog is at at http://astrogeo.org/petrov/.}
\tablenotetext{c}{Positions are measured from the images made from
  phase-referenced data that have not been self-calibrated.  The
  conservative error estimates include measurement errors, systematic
  phase-referencing errors and the errors in the position of the phase
  reference calibrator.}
\tablenotetext{d}{Phase-referencing check.  The astrometric catalog
  position from Petrov rfc\_2014a is R.A.=15 07 21.758075 ($\pm$ 0.13
  mas) and Decl. = 12 36 29.07581 ($\pm$ 0.21 mas).  It differs from
  our measured position by 0.18 mas in R.A. and 0.08 mas in Decl.}
\end{deluxetable}

A total of 256~MHz per circular polarization centered on $\nu$ =
4.980~GHz (5~GHz hereafter) were recorded using the new RDBE/MARK5C
wide-band system at 2 Gbps.  Every 1~s the VLBA DiFX correlator
\citep{del11} produced 512 contiguous 0.5-MHz channels.  About 3.4
hours were accrued on J1502+1115, during which the VLA position for
J1502S (F11) was used for pointing and correlation.  Correlation
parameters were chosen to ensure distortion-free imaging at the
position of J1502P, offset by 1.4\arcsec\, from J1502S (F11).  VLBA
system temperatures and gains were recorded for amplitude calibration.

Observations of OQ\,208 on 2013 July 1 UT were retrieved from the
archive of the Jansky Very Large Array (VLA) \citep[VLA;][]{per11}.
Release 4.1 of the Common Astronomy Software Applications (CASA)
package \citep{mcm07} was employed to calibrate and edit the data in
an automated fashion
\footnote{science.nrao.edu/facilities/vla/data-processing/pipeline}.
The final edits and imaging were done in the Astronomical Image
Processing System \citep[AIPS;][]{gre03}.  To best match the VLBA
data, the final VLA image of OQ\,208 utilized a total bandwidth of
128~MHz per circular polarization centered at $\nu$ = 5~GHz.  Direct
comparison with the primary flux calibrator 3C\,286 yielded a VLA flux
density of $2.33\pm0.07$~Jy, with the error dominated by an estimated
3\% uncertainty in the amplitude scale \citep{per13}.

AIPS was used for VLBA calibration and imaging.  The calibration
strategies documented in Appendix C of the AIPS Cookbook were
followed, including doing a bandpass calibration and making
corrections for the ionosphere and for updated Earth orientation
parameters.  The phase calibrator J1504+1029 was imaged using standard
methods based on multiple iterations of self-calibration and imaging.
Its final image was then used to determine the phase and amplitude
corrections to apply to all other data.  For OQ\,208, only the
amplitudes were of interest.  For J1507+1236, J1502S, and J1502P, the
phases enabled phase-referenced imaging, and those images were used to
measure the positions in Table~1.  After exploring various imaging
schemes, a robustness of 1 as implemented in AIPS was adopted.

The OQ\,208 data were self-calibrated and imaged, yielding an
integrated VLBA flux density of $1.78$~Jy, 31\% lower than the VLA
value.  This finding corroborated a recent report of low calibrated
flux densities from another VLBA user group.  The cause is under
investigation and seems to affect mainly the Polyphase Filter Bank
personality of the RDBE system when the bandpass is calibrated in what
is believed to be the proper manner.  Meanwhile, for this paper, we
have scaled the VLBA amplitudes upward by a factor of 1.31 to align
the VLA and VLBA photometry for OQ\,208, thereby calibrating the VLBA
amplitude scale to an accuracy of about 5\%.

The J1507+1236 data were self-calibrated and imaged, after which the
peak flux density increased by 78\% while the integrated flux density
did not change significantly, as expected when the emission is
concentrated by the improved phases.  Compared to 1502+1115, many
fewer observations were made of J1507+1236, and those observations
involved a switching angle about 2.5 times larger (Table~1).  Thus the
quality of the J1502+1115 data is expected to be much higher.  From
the self-calibrated data for J1504+1029 and J1507+1236, the integrated
flux densities are $990\pm50$~mJy and $208\pm10$~mJy, respectively,
with errors dominated by the 5\% uncertainty in the amplitude scale.

The J1502+1115 data contains information on J1502P and J1502S.  D14
did not detect J1502P, whereas our phase-referenced image of it
detects one radio source.  Our phase-referenced image of J1502S shows
that it consists of two radio sources, as reported by D14, with a
summed flux density of less than 2 mJy.  This image was used to
self-calibrate the phases of the J1502+1115 data.  Because J1502S is
so faint, the phase-only self-calibration was based on data coherently
averaged for 35~m over all frequencies and both polarizations.  This
long coherent average was possible because the initial phase
referencing had already removed the short-term phase fluctuations.
This phase self-calibration was used to reduce residual systematic
calibration offsets.  For J1502S, the improved calibration recovered
an additional $\sim$ 10\% in the peak flux densities and $\sim$ 5\% in
the integrated flux densities.  Self-calibration at low
signal-to-noise (S/N) can adjust the flux density upwards excessively
by gathering noise, or downward by failing to calibrate the residual
phase fluctuations.  To account for these effects, our experiments
with several coherent averages lead us to include an estimate of a 3\%
uncertainty in the error budget for the integrated flux densities
(term (ii) in Table~1).

Figure~1 shows the self-calibrated VLBA image of J1502S.  The achieved
rms noise is in line with the estimated thermal noise.  Figure~1
reveals that J1502S consists of two radio sources, labeled J1502SW and
J1502SE.  This image was used for the photometry in Table~1.  With
adequate S/N, structures as large as about 20 mas could be represented
in Figure~1.

\begin{figure}
\plotone{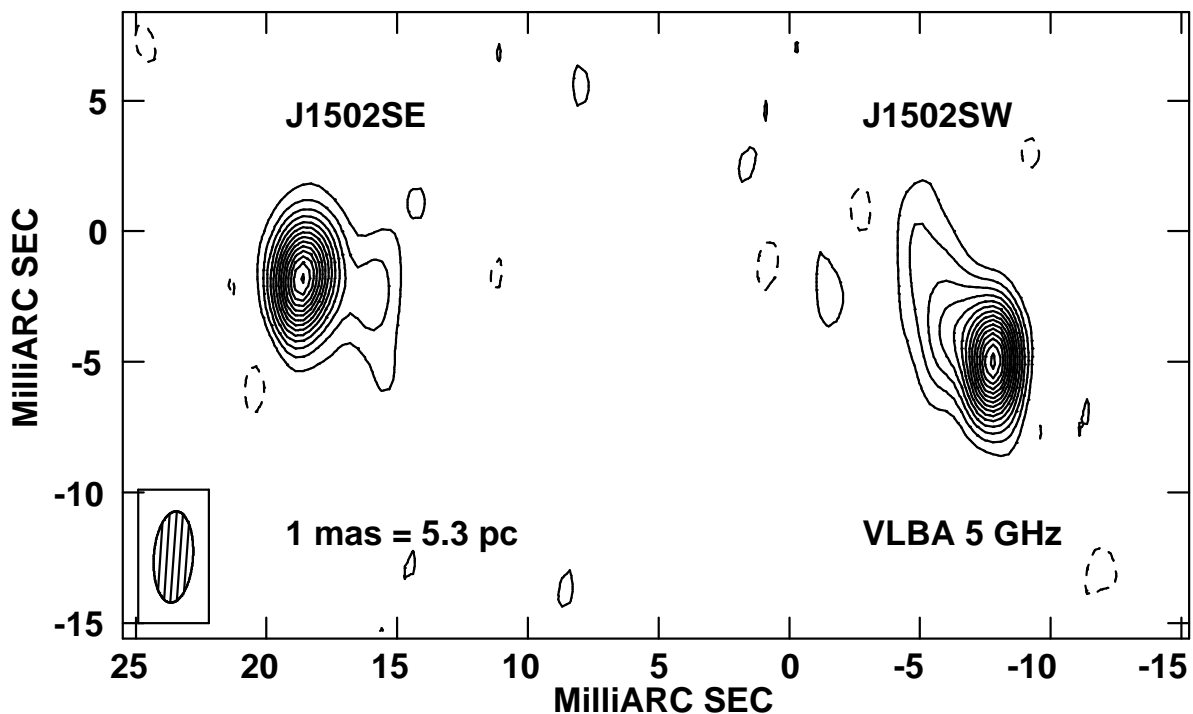}
\caption{VLBA image of Stokes $I\/$ emission from the dust-obscured
  AGN J1502S at $\nu$ = 5 GHz after self calibration.  The rms noise
  is $\sigma$ = 21~$\mu$Jy~beam$^{-1}$ and the synthesized beam
  dimensions at FWHM are 3.5 mas $\times$ 1.5 mas with an elongation
  position angle PA = $-$4\arcdeg\, (boxed hatched ellipse).  Allowed
  contours are at $-$6, $-$4, $-$2, 2, 4, 6, 8, 10, ... and 30 times 1
  $\sigma$.  Negative contours are dashed and positive ones are solid.
  The coordinate origin is at the VLA position for J1502S (F11).
  Clear edge-brightening is observed, suggesting that the two sources
  are hotspots energized by a single, centrally-located BH.  The
  western extension of J1502SE has a peak of 5.7
  $\sigma$.}\label{fig1}
\end{figure}

J1502P was imaged twice from the J1502+1115 data, once in its
phase-referenced form and once in its self-calibrated form.  The data
used for imaging retained the 0.5-MHz channels from the correlator to
prevent bandwidth smearing, and had its phase center shifted by
$-$1.329\arcsec\, in right ascension and 0.330\arcsec\, in
declination.  J1502P was detected as a slightly resolved source in
both images, with the self-calibration providing a 9\% enhancement in
the peak flux density and no change in the integrated flux density.
The signal from J1502P did not enter into the self-calibration, so
this is an independent check of the quality of that calibration.  An
elliptical Gaussian fit in the image plane was consistent with J1502P
being unresolved with major and minor axes less than 3.5 mas and 1.5
mas, respectively.  Since J1502P is so point-like, no image is
presented.

\section{Implications}\label{implications}

\subsection{J1502SE/SW}\label{j1502s}

From EVN images at 1.7 and 5~GHz, D14 reported evidence that the
dust-obscured AGN exhibits two flat-spectrum radio sources,
J1502SE/SW, offset by 26 mas (140 pc), with each source being
energized by its own supermassive BH.  While intriguing, this
interpretation of a close binary BH hole was adopted only after D14
had ruled out a more prosaic double-hotspot scenario.  In a
double-hotspot scenario both hotspots are energized by a single,
central BH.  Such a configuration occurs in a well-studied class of
radio-selected AGN, the CSOs \citep[][and references therein]{an12}.
CSOs have radio extents of less than 1 kpc; slow-moving emission on
both sides of the central engine; and weak, if any, radio variability
\citep[e.g.,][]{fas01}.  When observed with sufficient sensitivity and
resolution, the overall structure of a CSO can appear edge-brightened,
that is, the emission is brightest at the CSO's extremities, called
hotspots, and becomes fainter closer to its central BH.  The radio
sources J1502SE/SW were only slightly resolved in the EVN images,
making it difficult to look for edge-brightening.  However, the VLBA
image in Figure 1 clearly shows the expected edge-brightened structure
for J1502SE/SW.  For this reason, we conclude that the double-hotspot
scenario should be reconsidered as a viable interpretation for
J1502SE/SW.

When compared to CSOs in the compilation of \citet{an12}, D14 noted
that J1502SE/SW has several unusual properties, including an
atypically low spectral power of $P_{\rm{1.7~GHz}} = 7\times10^{23}$ W
Hz$^{-1}$ and an atypically flat spectral index of
$\alpha^{\rm{1.7~GHz}}_{\rm{5~GHz}} \sim -0.1\pm0.1$, especially given
its projected linear size.  These properties provide important clues
as to the nature of J1502SW/SE.  Below, we discuss these properties
within an alternate, and yet related, context that stems from two
findings for AGNs with radio powers similar to J1502SE/SW.

First, the radio source J1148+5924 has a spectral power of
$P_{\rm{1.4~GHz}} \sim 2\times10^{23}$ W Hz$^{-1}$, a factor of a few
below that of J1502S \citep{tay98}.  J1148+5924 is dominated by radio
emission with an extent of less than 1 kpc, shows a slow apparent
separation speed of 0.3c on parsec scales, and displays slow radio
variability \citep{tay98,pec00,fas01}.  Such traits are shared by the
powerful CSOs \citep{an12}, suggesting that J1148+5924 is a low-power
CSO.  Moreover, J1502S shows evidence for an overall rotational
symmetry on scales 0.1-10 kpc (D14).  J1148+5924 also shows evidence
for an overall rotational symmetry on similar scales, possibly caused
by interactions between the low-power radio outflow and the
circumnuclear gas and dust \citep{wro84,tay98,per01}.  Redshifted HI
is seen in absorption against the two-sided structures in J1148+5924
\citep{pec98}, and this infalling gas could fuel and/or distort the
slow-moving radio outflow.

Second, the obscured AGN J1502S was originally [OIII]-selected
\citep{smi10} and has a luminosity of $L({\rm [OIII]}) = 3.6 \times
10^{42}$ erg s$^{-1}$.  In their analysis of the radio properties of
obscured AGN that have $L({\rm [OIII]}) \sim 10^{42}$ erg s$^{-1}$,
\citet{lal10} find a high incidence of spectra that are flat or rising
between 1.4 and 8.4 GHz.  Those authors speculate that such radio
spectra could be caused by free-free absorption in the ionized gas
traced in [OIII].  J1502S has a similar [OIII] luminosity.  This hints
that free-free absorption could contribute to the flat spectra D14
measure for J1502SE/SW, although achieving flatness between 1.7-15.7
GHz could be a challenge for such an absorption model.  Still, this
suggestion is testable with VLBA imaging spanning a wide frequency
range \citep[e.g.,][]{mar14}, which could also help reveal emission
from a centrally located origin of activity.

D14 cite a 10\% uncertainty in the amplitude scale of the EVN at 5
GHz.  Factoring this into their reported photometry leads to flux
densities of 857$\pm$99 $\mu$Jy and 872$\pm$100 $\mu$Jy for J1502SE
and J1502SW, respectively, in 2011 April 12.  Comparison with the VLBA
photometry (Table 1) obtained more than two years later implies no
time variability, another trait consistent with CSO-like behavior.
From the EVN images, the apparent separation between J1502SE and
J1502SW increases by less than 5 pc yr$^{-1}$.  For comparson, for
J1148+5924 the apparent separation increases by 0.1 pc yr$^{-1}$.  If
J1502SE/SW have a similarly-slow apparent separation, it would be
advantageous to monitor that separation with the VLBA's higher
sensitivity and higher resolution.

More broadly, establishing a double-hotspot scenario for the
dust-obscured AGN J1502S could have important implications for
feedback in obscured AGNs, whether those AGN are discovered in surveys
for X-ray or infrared continuum or for narrow optical emission lines
\citep[][and references therein]{lal10} or in increasingly deep VLBA
surveys \citep[][and references therein]{mid13,del14}.  Conversely,
our study underscores the importance of culling double-hotspot sources
from VLBA surveys seeking candidate binary BHs on parsec scales
\citep[e.g.,][]{bur11,tin11} or strong gravitational lenses on mas
scales \citep{wil01}.

\subsection{J1502P}\label{j1502p}

From Table~1 the VLBA detection of J1502P at $\nu$ = 5 GHz has a
rest-frame brightness temperature of 4.8 million K, or more if truly
unresolved.  Such levels are not achieved by even the most compact
startbursts \citep{con92}.  Also, the VLBA image is too shallow to
detect even the most luminous radio supernovae beyong $z \sim 0.1$
\cite[e.g.,][]{mid13}.  Thus the VLBA detection indicates that J1502P
must be AGN driven, confirming, independent of F11, that J1502P hosts
an AGN.  At $\nu$ = 5 GHz, the VLA photometry (F11) localizes the
emission to a diameter of 300 mas (1.6 kpc), whereas the EVN
photometry (D14), obtained only about two months earlier, recovers
less than 20\% of the VLA signal.  Both this fact and the overall
steep-spectrum nature of J1502P (F11) suggest that source-resolution
effects are at play, a motivation for deeper VLBA imaging.

\section{Summary and Conclusions}\label{summary}

We used the VLBA at 5~GHz to image J1502+1115, a merging system at $z
= 0.39$ containing a dust-obscured AGN, J1502S, offset by 1.4\arcsec\,
(7.4 kpc) from an unobscured AGN, J1502P (F11).

Regarding J1502S, D14 advocate it as hosting two AGNs and, thus, two
supermassive BHs, based on EVN images at 1.7 and 5~GHz showing two
slightly-resolved, flat-spectrum radio sources, J1502SE/SW, offset by
26 mas (140 pc).  D14 reached their intriguing interpretation of a
140-pc binary BH after discounting a double-hotspot scenario, wherein
both hotspots are energized by a single, central BH, a configuration
occuring amongst radio-selected CSOs.  When observed with sufficient
sensitivity and resolution, an object with double hotspots should have
an edge-brightened appearance.  We find clear evidence for such
edge-brightening in our VLBA image of J1502S that has higher
sensitivity and resolution than the EVN images.  We thus conclude that
the double-hotspot scenario should be reconsidered as a viable
interpretation for J1502SE/SW.  We also suggest that free-free
absorption by J1502S's [OIII] emitting gas could help flatten the
radio spectra of J1502SE/SW.  Future VLBA imaging can further test the
double-hotspot scenario for J1502SE/SW, as well as the potential role
of free-free absorption.  A double-hotspot scenario could have
important consequences for feedback modes in obscured AGNs, a key
population for understanding evolutionary linkages between galaxies
and the BHs they host. Future VLBA imaging of samples of low-power
CSOs is also needed to investigate trends between spectral indices and
projected linear sizes, to help distinguish low-power CSOs from close
binary BHs.  And because the detection of polarized intensity is
relatively rare among powerful CSOs \citep{hel07}, such VLBA imaging
should include polarimetry as a possible discriminant between
low-power CSOs and close binary BHs.

Regarding J1502P, as it is detected in our VLBA image, it must be AGN
driven.  This confirms, independent of F11, that J1502P hosts an AGN.
The VLBA detection is faint, only 254 $\mu$Jy.  Future, deeper VLBA
imaging is needed to characterize it further.

\acknowledgements We thank the anonymous referee and the commentator,
Roger Deane, for prompt and helpful feedback.  This work made use of
the Swinburne University of Technology software correlator, developed
as part of the Australian Major National Research Facilities Programme
and operated under licence.  We acknowledge using Ned Wright's
Cosmology Calculator \citep{wri06}.  We are grateful to Drew Medlin
for providing only the calibrator scans from the VLA data archived on
2013 July 1.

{\it Facilities:} \facility{VLA}, \facility{VLBA}.

\end{document}